\documentclass[%
 reprint,
 amsmath,amssymb,
 aps,
]{revtex4-2}

\usepackage{graphicx}
\usepackage{dcolumn}
\usepackage{bm}
\usepackage{color,soul}
\usepackage{fancyhdr}
\usepackage{bm}
\usepackage{psfrag}
\usepackage{graphicx}
\usepackage{dcolumn}
\usepackage{bm}
\usepackage{todonotes}
\usepackage{hyperref}
\usepackage[utf8]{inputenc}
\usepackage[T1]{fontenc}
\usepackage{mathptmx}
\usepackage{etoolbox}
\usepackage{amsmath}
\usepackage[utf8]{inputenc}
\usepackage{float}

\definecolor{ForestG}{rgb}{0.31, 0.78, 0.47}
\definecolor{MyGreen}{RGB}{54,165,54}

\makeatletter
\def\@email#1#2{%
 \endgroup
 \patchcmd{\titleblock@produce}
  {\frontmatter@RRAPformat}
  {\frontmatter@RRAPformat{\produce@RRAP{*#1\href{mailto:#2}{#2}}}\frontmatter@RRAPformat}
  {}{}
}%
\makeatother
\begin{document}

\preprint{APS/123-QED}

\title{Theoretical study of the impact of carrier density screening on Urbach tail energies and optical polarization in (Al,Ga)N quantum well systems}

\author{Robert Finn$^{1}$} 
\thanks{Contact author: robert.finn@tyndall.ie}%
\author{Michael O'Donovan$^2$}
\author{Thomas  Koprucki$^2$}
\author{Stefan Schulz$^{1,3}$}%
\affiliation{$^1$Tyndall National Institute, University College Cork,
Cork, T12 R5CP, Ireland}
 \affiliation{$^2$Weierstrass Institute (WIAS), Mohrenstr. 39, 10117 Berlin, Germany}
\affiliation{$^3$School of Physics, University College Cork, Cork, T12 YN60, Ireland}%

\date{\today}

\begin{abstract}

Aluminium Gallium Nitride ((Al,Ga)N) presents an ideal platform for designing ultra-violet (UV) light emitters across the entire UV spectral range. However, in the deep-UV spectral range (<280 nm) these emitters exhibit very low quantum efficiencies, which in part is linked to the light polarization characteristics of (Al,Ga)N quantum wells (QWs). In this study we provide insight into the degree of optical polarization of (Al,Ga)N QW systems operating across the UV-C spectral range by means of an atomistic, multi-band electronic structure model. Our model not only captures the difference in valence band ordering in AlN and GaN, it accounts also for alloy disorder induced band mixing effects originating from random alloy fluctuations in (Al,Ga)N QWs. The latter aspect is not captured in widely employed continuum based models. The impact of alloy disorder on the electronic structure is studied in terms of Urbach tail energies, which reflect the broadening of the valence band density of states due to carrier localization effects. We find that especially in wider wells, Urbach tail energies are reduced with increasing carrier densities in the well, highlighting that alloy disorder induced carrier localization effects in (Al,Ga)N QWs are also tightly linked to electrostatic built-in fields. Our calculations show that for QWs designed to emit at the longer wavelength end of the UV-C spectrum, carrier density and well width are of secondary importance for their light emission properties, meaning that one observes mainly transverse electrical polarization. However, for (Al,Ga)N QWs with high Al contents, we find that both well width and carrier density will impact the degree of optical polarization. Our calculations suggest that wider wells will increase the degree of optical polarization and may therefore be a viable option to improve the light extraction efficiency in deep UV light emitters. 
\end{abstract}

\maketitle

\section{Introduction}
The semiconductor alloy aluminium gallium nitride, (Al,Ga)N, has gained significant attention in the last decade as the direct band gap of this semiconductor can in principle span 
the entire UV range (UV-A (315-400 nm), UV-B (280-315 nm) and UV-C (100-280 nm)). 
Applications that require efficient UV light sources, e.g. light emitting 
diodes (LED) or lasers, operating over this wide spectral window include water purification, sterilization, plant lighting, sensing, etc~\cite{AmCo2020_JPD}. (Al,Ga)N-based UV light emitting devices come with significant advantages over widely employed low and medium pressure mercury lamps as they (i) provide flexible and tuneable wavelengths, (ii) do not require toxic mercury, (iii) show no warm up-time and (iv) exhibit  extremely long lifetimes~\cite{AmCo2020_JPD, KnSe2019}.  
However, when compared to indium gallium nitride, (In,Ga)N, a light emitter operating in the visible spectral range, current UV light emitting devices utilizing (Al,Ga)N alloys still exhibit poor external quantum efficiencies especially in the deep UV-range (<280 nm). Multiple factors contribute to the low efficiency of e.g. (Al,Ga)N LEDs which include high threading dislocations and point defect densities, poor radiative recombination rates and low light extraction efficiencies (LEE)~\cite{AmCo2020_JPD}.
Especially the latter two aspects are tightly linked to fundamental properties of (Al,Ga)N alloys and heterostructures. Firstly, the thermodynamical stable phase of (Al,Ga)N alloys is the wurtzite crystal structure~\cite{AmCo2020_JPD,ChCh1996}, and its lack of inversion symmetry results in a spontaneous polarization field along the wurtzite $c$-axis. In a nitride heterostructure, e.g. an Al$_x$Ga$_{1-x}$N/Al$_y$Ga$_{1-y}$N quantum well (QW) which lies at the heart of UV LEDs, discontinuities in spontaneous polarization lead to strong electrostatic built-in fields~\cite{AmCo2020_JPD,CaSc2011,AmMA2002}. In addition, the lattice mismatch between Al$_x$Ga$_{1-x}$N/Al$_y$Ga$_{1-y}$N gives rise to a strain induced piezoelectric polarization field~\cite{BeFi1997}. These electrostatic built-in fields result in a spatial separation of electron and hole wave functions in a QW, affecting and reducing the radiative recombination rate~\cite{Re2016}. Moreover, the LEE is impacted by the valence band ordering in an (Al,Ga)N heterostructure. In general, while in both GaN and AlN the conduction band minimum and the valence band maximum are at the $\Gamma$-point, the valence band structure of AlN is different from its GaN counterpart~\cite{GoBu2007,SaBe2010,BaFu2009,WeZu1996}. More specifically, 
the valence band edge (VBE) in AlN is formed by a band of $\Gamma_{7+}$ symmetry, so that the radiative recombination process involving this band and the conduction band at the $\Gamma$-point results in light that is mainly of transverse magnetic (TM) polarization. In GaN the VBE is of $\Gamma_{9}$ character; in this case the emitted light is mainly transverse electric (TE) polarized. As LED structures are in general surface emitting, the TM polarized light emission is detrimental for the LEE and thus the efficiency of the device~\cite{ShCh2022,KnKo2010,NoCh2012,RyCh2013,Ry2014}. The Al content in the well, but also quantum confinement and strain effects affect the crossover from TE to TM polarization in (Al,Ga)N-based LEDs. In addition to these factors, experimental and theoretical studies already give indications that alloy-disorder induced carrier localization effects significantly impact the optical properties of (Al,Ga)N QWs~\cite{FrNi2020_JAP,FrNi2020_PSS,FiSc2022,KoRo2024}. Carrier localization can manifest for instance through broad photoluminescence (PL) linewidth or broad absorption spectra, where the latter is often described in terms of Urbach tail energies~\cite{PiLi2017,McMTa2020,RoGu2021}.

The impact of alloy disorder on the degree of optical polarization (DOP), usually defined as the ratio of TE to TM polarized emission~\cite{NaLi2004,BaFu2009}, is largely unexplored in theoretical studies of (Al,Ga)N QWs as it presents a significant challenge for several reasons. Firstly, standard and widely available one dimensional multi-band $\mathbf{k}\cdot\mathbf{p}$ simulations of (Al,Ga)N QWs and heterostructures in general describe the alloy by averaged material parameters; local fluctuations in the alloy content are in general neglected~\cite{JaSu2017,HoMu2024}. Secondly, modified continuum based models, which allow for local fluctuations in alloy content in a three dimensional simulation, have recently been employed to describe the impact of alloy disorder on the electronic and optical properties of (In,Ga)N QWs; such studies are in general based on single-band effective mass models~\cite{YaSh2014,WaGo2011,ChODo2021}. These models cannot capture intrinsically any alloy disorder induced valence band mixing effects. Thirdly, to treat carrier localization effects atomistic calculations are in principle an ideal starting point. However, to do so large (in-plane and out-of-plane) supercells are required to capture both disorder in the growth plane but also barrier materials. First principles approaches, such as density functional theory (DFT), can provide the required microscopic insight into the electronic structure of III-N materials and alloys. However, in general it requires very large supercells to capture carrier localization effects accurately. For `conventional' DFT implementations this presents a significant, basically unfeasible computational challenge, especially when dealing with QW heterostructures~\cite{BrOr2024,CaSc2012}. 

In recent years, empirical tight binding models (ETBMs) have demonstrated that they can capture carrier localization effects accurately on an atomistic level while at the same time allowing to treat very large supercells~\cite{BrOr2024}. We have developed ETBMs to describe the electronic structure of (Al,Ga)N bulk and QW systems~\cite{CoSc2015,FiSc2022}. Here, we employ this atomistic multi-band model to gain insight into Urbach tail energies and the DOP of (Al,Ga)N-based QWs. Given that alloy composition and quantum confinement on a macroscopic level will play an important role for the electronic structure of (Al,Ga)N wells in general, we investigate QWs with two distinctly different alloy compositions (Al$_{0.48}$Ga$_{0.52}$N/Al$_{0.63}$Ga$_{0.37}$N and Al$_{0.75}$Ga$_{0.25}$N/Al$_{0.90}$Ga$_{0.10}$N) and of varied well widths ($L_w$=1.3 nm, 2.3 nm and 3.3 nm). Al$_{0.48}$Ga$_{0.52}$N/Al$_{0.63}$Ga$_{0.37}$N wells often form the active region of UV-C emitters operating at the longer wavelength end of the UV-C spectrum~\cite{RoGu2021,SuZi2020}. Based on literature data, these emitters are expected to mainly emit TE polarized light. On the other hand, Al$_{0.75}$Ga$_{0.25}$N/Al$_{0.90}$Ga$_{0.10}$N QWs were chosen to investigate a system emitting (i) in the deep UV range and (ii) mainly TM polarized light; similar Al contents are reported in the literature for deep UV emitters~\cite{RoGu2021,RoHo2024IEEE,IbLe2024}. It should be noted that both systems exhibit the same 15\% Al composition contrast between well and barrier. Thus, our analysis should allow for a comparative investigation between emitters operating in two distinct UV-C wavelength ranges. Our studies also target varying carrier densities in the well in order to gain insight into experimental relevant settings.

Our atomistic multi-band investigations reveal that carrier confinement, especially for narrower wells, e.g. \mbox{$L_w=1.3$ nm}, is much weaker in the high Al content system Al$_{0.75}$Ga$_{0.25}$N/Al$_{0.90}$Ga$_{0.10}$N when compared to the Al$_{0.48}$Ga$_{0.52}$N/Al$_{0.63}$Ga$_{0.37}$N QW structures. While this effect may be beneficial for carrier transport in multi QW structures operating in the deep UV spectral region, the connected carrier leakage may also lead to an increase in defect related recombination. 

When studying the DOP of the QW systems considered, we find that wider wells lead to a higher DOP value, which in general is desired for good light extraction efficiencies for UV LEDs. However, in Al$_{0.48}$Ga$_{0.52}$N/Al$_{0.63}$Ga$_{0.37}$N wells this finding is of secondary importance as the light emitted for all well widths and carrier densities studied is already predominately of transverse electric (TE) polarization. For the higher Al content structure, the Al$_{0.75}$Ga$_{0.25}$N/Al$_{0.90}$Ga$_{0.10}$N wells, the situation is more subtle. While the DOP increases with increasing well width, it is in general still transverse magentic (TM) polarized.  
We observe that the shift towards TE polarization with increasing well width is stronger in the lower carrier density regime, e.g. $n=1\times10^{18}$ cm$^{-3}$. 
Nevertheless, even for a high carrier density of $n=1\times10^{20}$ cm$^{-3}$, an Al$_{0.75}$Ga$_{0.25}$N/Al$_{0.90}$Ga$_{0.10}$N QW of width $L_w=3.3$ nm still exhibits a slightly higher DOP when compared to the same system with a width of $L_w=1.3$ nm. This again highlights the potential benefit of wider wells to improve and tailor the LEE in deep UV (Al,Ga)N based emitters. 

The manuscript is organised as follows. In Sec.~\ref{sec:theory} we discuss the theoretical background of our studies, while Sec.~\ref{sec:results} presents our results. Finally, summary and conclusion is given in Sec.~\ref{sec:Concl}.

\section{Theoretical Background}
\label{sec:theory}

In Sec.~\ref{sec:VB} we provide an overview of differences in the valence band structure of AlN and GaN systems which are important for understanding the DOP in (Al,Ga)N based QWs. The electronic structure model underlying our investigations of the DOS and DOP is outlined in Sec.~\ref{sec:TBmodel}; the (Al,Ga)N QWs studied in this work are discussed in Sec.~\ref{sec:QWModel}. 

\subsection{Valence Band Structure of AlN and GaN}
\label{sec:VB}
The DOP in (Al,Ga)N alloys is tightly linked to the symmetry properties of the valence bands involved in the optical transitions between conduction and valence bands. Given the direct band gap 
nature of both wurtzite GaN and AlN, the symmetry of the three valence bands energetically closest to the band edge at the $\Gamma$-point 
play an important role for the DOP; lower lying valence states are energetically far away from the VBE at $\Gamma$. In general, the symmetry of the energetically highest lying three valence bands can be described by $p_x$-, $p_y$- and $p_z$-like orbitals or respective linear 
combinations~\cite{ChCh1996}. Spin-orbit coupling and crystal field splitting lift the degeneracy between these bands, which are sometimes termed as $A$, $B$ and $C$ valence 
bands~\cite{WiSc2018,RiWi2008,NaLi2004}. However, the ordering of these bands depends on several factors. For instance, the crystal field splitting energy, $\Delta_\text{cr}$, in AlN is of \emph{opposite} sign to $\Delta_\text{cr}$ in GaN or InN. In the literature, $\Delta_\text{cr}$ values can vary ~\cite{YaRi2011,VuMe2003}, but for AlN $\Delta^\text{AlN}_\text{cr}\approx-200$ meV whereas for GaN $\Delta^\text{GaN}_\text{cr}\approx30$ meV. The sign of $\Delta_\text{cr}$ determines which valence band forms the VBE as the spin-orbit coupling energy is usually small, particularly when compared to other III-V material systems~\cite{VuMe2001}.

The polarization state of the light emitted via the radiative recombination process involving the conduction band edge (CBE) and VBE is determined by the symmetry (orbital character) of the VBE state at the $\Gamma$-point. Due to the negative $\Delta^\text{AlN}_\text{cr}$ value in AlN, the VBE is  formed by the so-called $C$-band which is of $\Gamma_{7+}$  symmetry~\cite{GoBu2007,SaBe2010,BaFu2009,NaLi2004}; this band is sometimes also called the crystal-field split off hole (CH) band. Energetically below the $C$-band are the $A$- and $B$-bands which are mainly split by the small spin-orbit coupling energy in nitrides ($\approx$ 10-20 meV)~\cite{VuMe2001}. The $A$-band is of $\Gamma_9$ symmetry and often labeled as the heavy hole (HH) band; the $B$-band, or light hole (LH) band, has $\Gamma_{7-}$ symmetry. In terms of the basis states forming these bands, the upper $\Gamma_{7+}$ band exhibits mainly $p_z$-like character while both $\Gamma_9$ and $\Gamma_{7-}$ are mainly $p_x$- and $p_y$-like in character~\cite{BaFu2009}. For GaN, due to the positive crystal field splitting energy, the valence band order is reversed: the VBE is the $A$-band; $B$- and $C$-bands are energetically below the VBE~\cite{RiWi2008,NaLi2004}. Given the difference in the orbital character/symmetry of the VBE, light emitted from AlN is TM polarized while in GaN it is of TE polarization~\cite{BaFu2009,ShCh2022}. The latter polarization leads to light emission mainly along the $c$-axis of the underlying wurtzite crystal. As standard UV (Al,Ga)N LEDs are top and bottom surface emitting and are grown along the wurtzite $c$-axis, TE polarized light emission is desired to achieve a high LEE of the device~\cite{BaFu2009,ShCh2022,KnKo2010}.  

The above highlights that the valence band ordering in (Al,Ga)N-based QWs plays an important role when designing UV LEDs, especially in the deep-UV, where high Al contents in the active region are required. However, in addition to the Al content, further factors impact the valence band ordering in Al$_x$Ga$_{1-x}$N/Al$_y$Ga$_{1-y}$N wells. 
These factors include contributions from strain~\cite{ShCh2022}, quantum confinement and built-in polarization fields~\cite{BaFu2009}.
 
\begin{figure}[t!]
\includegraphics[width=\columnwidth]{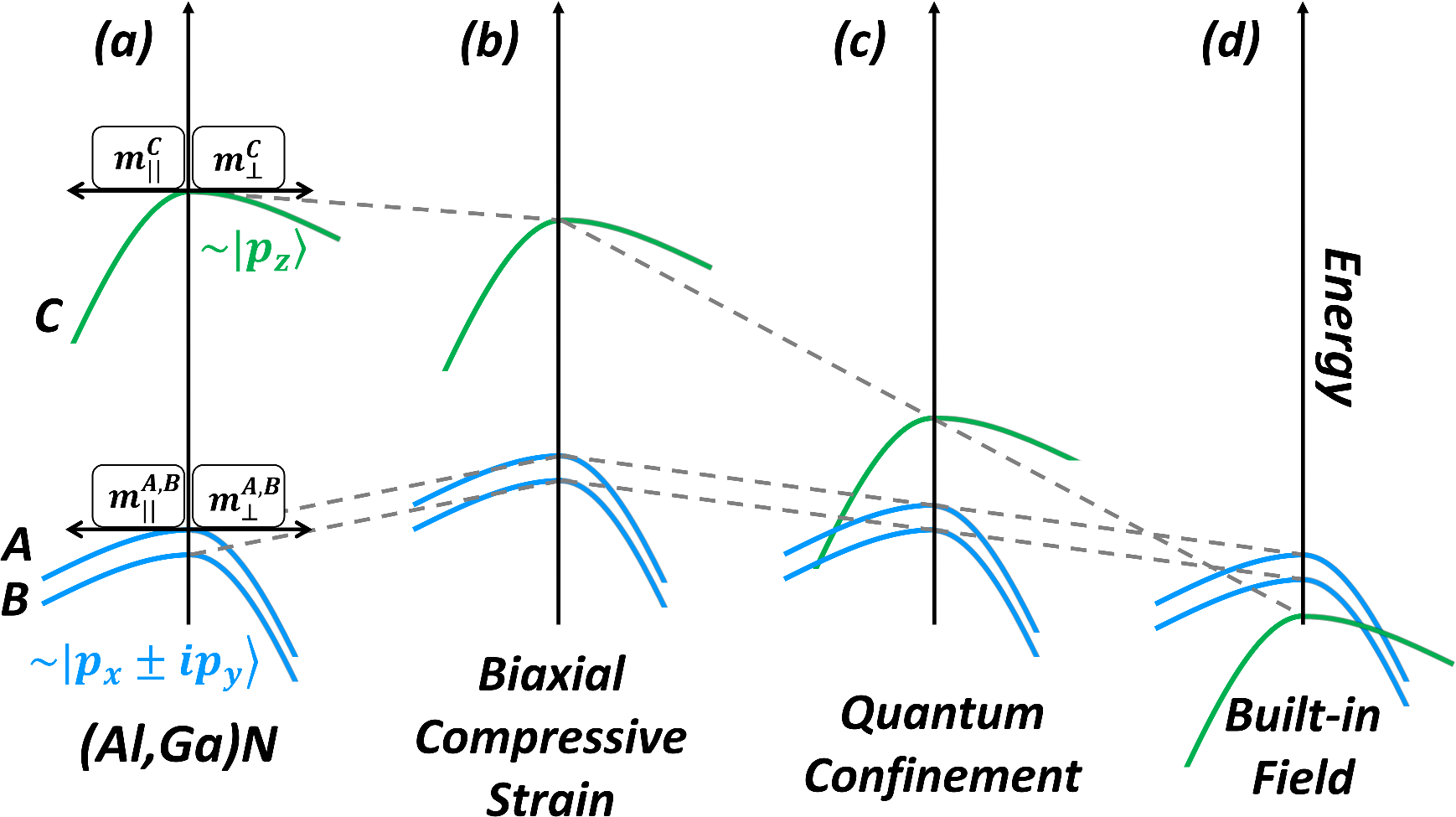}
\caption{Schematic illustration of the valence band order for high aluminium containing (Al,Ga)N systems when (a) unstrained, (b) under biaxial compressive strain, (c) in then presence of quantum confinement and (d) intrinsic built-in fields. The heavy hole, light hole and crystal field split-off bands are denoted as the $A$, $B$ and $C$ bands, respectively. Note that our schematic is to highlight general trends. Quantum confinement will break the translation invariance e.g. along the wurtzite $c$-axis and a band structure cannot be defined along this direction. The above should however indicate how quantized energy levels that are constructed from e.g. $|p_z\rangle$-like basis states are affected by confinement effects.}
\label{fig:band_schem}
\end{figure}

Figure~\ref{fig:band_schem} gives a schematic illustration of how strain, confinement and built-in field affect the valence band ordering in an (Al,Ga)N system. In Fig~\ref{fig:band_schem}~(a) a high Al content is assumed so that the VBE is given by the $C$-band ($p_z$-like band) and the band ordering is illustrated in the absence of strain, confinement and built-in polarization field effects. The impact of in-plane biaxial compressive strain on the band structure, as for instance found in a QW structure, is illustrated in Fig~\ref{fig:band_schem}~(b). With respect to the unstrained case, $A$ and $B$ bands are shifted to higher energies while the $C$ band shifts to lower energy. Depending on strain and Al content in the well, this may now already lead to a crossover of the $C$- and $A$/$B$ bands, but in any case a reduced energy separation between the $C$ band ($p_z$-like states) and $A$ and $B$ bands (mainly $p_x$- and $p_y$-like states) is expected. The reduced energetic separation between the different bands can thus lead to band mixing effects (e.g. mixing of $p_x$-, $p_y$- and $p_z$-like states) which affect the DOP.

Figure~\ref{fig:band_schem}~(c) visualises the impact of quantum confinement along the $c$-axis (standard growth direction for UV LEDs) on the band separation. This type of confinement further reduces the energetic separation of the three bands in our example. This stems from the fact that the effective hole mass of the $C$-band along the confinement/growth direction is lower when compared to the $A$- and $B$-bands; for instance in AlN $m^{h,C}_{||}=0.25m_0$ and $m^{h,A}_{||}=3.125m_0$~\cite{RiWi2008} where $m_{||}$ denotes the effective mass parallel to the $c$-axis. Therefore, to tailor the valence band ordering and thus light polarization characteristics, one may adjust the QW width to modify and control the quantum confinement. However, this may also be achieved by changing the contrast in Al composition between well and barrier since this will modify the valence band offset.

A further factor that affects the confinement in a III-N well is the built-in polarization field. This field spatially separates electron and hole wave functions by localizing them at opposite interfaces of the well. Thus, the hole wave function may be confined to a length smaller than the well width $L_w$; this ``additional'' confinement can further affect the valence band ordering as schematically indicated in Fig.~\ref{fig:band_schem}~(d). Several factors impact the strength of this field including the composition difference between QW and barriers and the overall strain in well and barrier. However, with an increasing carrier density in the QW, the polarization field may be screened so that this polarization field induced confinement effect is reduced. This means that the valence band ordering in a (Al,Ga)N QW system can also be carrier density dependent.

The above is based on the assumption that in an (Al,Ga)N QW all these different contributions can be described on a macroscopic level, i.e. that local features of the alloy microstructure are of secondary importance. Our recent calculations~\cite{FiSc2022} but also experimental data~\cite{FrNi2020_JAP,FrNi2020_APL} indicate that alloy disorder can give rise to carrier localization effects in (Al,Ga)N QWs. Such localization effects can be driven by local strain, confinement and polarization field fluctuations. The importance of these features for the DOP are not well understood and require ideally an atomistic electronic structure model that also captures the different valence bands in (Al,Ga)N QWs. Below we outline our theoretical framework that meets these criteria.

\subsection{Empirical Tight-Binding Model}
\label{sec:TBmodel}

As already mentioned in the previous section, alloy disorder can influence the electronic and optical properties of (Al,Ga)N QWs significantly. To address the impact of alloy disorder on the electronic structure and ultimately the DOP on an atomistic level, we employ a nearest neighbour multi-band $sp^3$ empirical tight-binding model (ETBM). The details of the model have been presented in our previous work in detail on bulk and QW systems~\cite{FiSc2022,CoSc2015}. The main ingredients involve a valence force field model to determine relaxed atomic positions in an (Al,Ga)N QW. To account for polarization field effects stemming from e.g. (local) strain fields we have developed a local polarization potential method. This method has been benchmarked against density functional theory based Berry phase calculations. The tight-binding parameters required for the $sp^3$ ETBM have been treated as parameters and determined by fitting the ETBM electronic band structure to hybrid functional DFT band structures. To model alloys with this approach, the ETBM hopping matrix elements are set according to nearest neighbour anion-cation species; for the on-site energies of Ga and Al atoms the respective bulk values are used. In case of N atoms, weighted averages of the on-site bulk values in AlN and GaN are employed; the weighted averaged are determined by the number of Ga and Al atoms as nearest neighbours of an N atom; this is a widely used approximation in ETBMs~\cite{OReLi2002,LiPo1992,BoKh2007}. A valence band offset of 0.9 eV for AlN/GaN is assumed, which lies within the range of reported values of 0.85-1.15 eV~\cite{MoMi2011,Mo1996JAP}. To account for strain effects, a local strain tensor is calculated and via a Pikus-Bir approach included as a site-diagonal correction in the ETBM Hamiltonian~\cite{CaSc2013local}. The required deformation potentials are again taken from hybrid functional DFT studies. Finally, the (local) built-in polarization potentials are included as an on-site diagonal correction to the Hamiltonian. Overall, our ETBM has been benchmarked against experimental data, showing a good agreement between the predicted energy gap of bulk Al$_x$Ga$_{1-x}$N alloys as a function of the Al composition $x$~\cite{CoSc2015}. Moreover, good agreement between the ETBM and experiment was found for the optical polarization switching in Al$_x$Ga$_{1-x}$N bulk alloys~\cite{CoSc2015}. Based on these previous studies, our established ETBM presents an ideal starting point to investigate the impact of alloy disorder on the DOP in (Al,Ga)N QW systems.

\subsection{Model Quantum Well System, Density of States and Degree of Optical Polarization}
\label{sec:QWModel}

Overall, our aim is to study the impact of alloy disorder on the electronic and optical properties, or more specifically the DOP, of (Al,Ga)N based QWs operating in different wavelength windows. In general, the emission wavelength of such structures is mainly tailored by adjusting Al content and well width. As discussed above, both factors will impact the valence band ordering and thus the DOP. We focus on two systems in the following, targeting the upper UV-C range ($\approx$ 280 nm) and deep UV-C emitters ($\approx$ 230 nm) to gain insight into similarities, but also differences in these wavelength ranges. 

The first QW system investigated is an Al$_{0.48}$Ga$_{0.52}$N well with an Al$_{0.63}$Ga$_{0.37}$N barrier. Such structures are often considered in the literature to realise emitters in the upper UV-C range~\cite{RoGu2021,SuZi2020}. Literature data shows that the considered system exhibits TE polarized light emission~\cite{RoGu2021,SuZi2020}. To also study the impact of alloy disorder on electronic and optical properties of (Al,Ga)N-based emitters in the deep UV-C range, we investigate Al$_{0.75}$Ga$_{0.25}$N QWs with Al$_{0.90}$Ga$_{0.10}$N barriers. For such structures, literature results find TM polarized light emission due to the high Al content in the well~\cite{IbLe2024,BaFu2009}. We note that while Al contents in the well and barrier are different for the two structures, the Al content \emph{contrast} between barrier and well is kept constant in our studies. Furthermore, we assume that the wells and barriers are strained to a pure AlN substrate. 

To study the impact of the well width, $L_w$, on the DOP we consider for both systems three different well widths, namely  $L_w\approx$1.3 nm, 2.3 nm and 3.3 nm. The electronic structure of the (Al,Ga)N well systems with the different well widths discussed above is modelled with our atomistic ETBM on supercells with dimensions of approximately $10\times9\times10$ nm$^3$ (81,920 atoms).

To gain insight into the DOP it is important to note that in an `ideal' QW structure, i.e. in the absence of carrier localization effects, the valence sub-band structure also plays an important role. At elevated temperatures or carrier densities, carriers may not only populate the topmost valence band but also energetically lower lying bands, which in (Al,Ga)N wells may exhibit different orbital and as such different light polarization characteristics. Thus, in general an orbital resolved density of states (DOS) can provide important insight into the DOP and how it changes with e.g. well width or Al content. However, in the presence of alloy-disorder induced carrier localization effects, defining valence sub-bands is in general not possible since the in-plane wave vector is no longer a good quantum number. Moreover, carrier localization effects lead to a broadening of the DOS, usually characterised by a so-called Urbach tail and a corresponding Urbach tail energy. Thus, it is useful to study changes in the Urbach tail energies for (Al,Ga)N QW systems before turning to the DOP.

In the following we focus on Urbach tail energies for the valence band as (i) our previous studies showed that alloy disorder effects impact hole states more significantly than electron states and (ii) the DOP is tightly linked to the orbital character of the valence/hole states. In order to do so we calculate the hole density of states (h-DOS) employing the atomistic ETBM introduced above. Given that hole state energies are significantly impacted by alloy disorder and to resolve especially the Urbach tail, for each of the (Al,Ga)N QW systems considered, 150 microscopic alloy configurations are generated per well width. To capture an energy range that is wide enough to resolve the DOS and especially the Urbach tail, 150 states for each well width were calculated per alloy configuration.

Furthermore, as already mentioned above, the intrinsic electrostatic built-in field in an (Al,Ga)N QW is screened with increasing carrier density in the well. As a consequence, the electronic structure of a well is also changed and as such impacts the Urbach tails and the DOP. In order to account for the polarization field screening effect, a self-consistent Schr\"odinger-Poisson calculation is required. However, performing a self-consistent ETBM-Poisson calculation on our large, atomistic 3D supercell for different well widths and (Al,Ga)N QW systems as well as varying carrier densities is numerically prohibitive. Instead, we employ the method established in Ref.~\onlinecite{McSc2024} where the screening potential at a given carrier density in the well is determined  by a self-consistent \mbox{1-D}  Schr{\"o}dinger-Poisson calculation, which is then included in the full atomistic ETBM. Diagonalising the `screened' ETBM Hamiltonian accounts for changes in the electronic structure due to built-in field screening effects, while still accounting for the impact of alloy disorder. Our recent studies on (In,Ga)N QW systems have shown that the method outlined produces results in good agreement with experiment in terms of e.g. carrier densities at which screening of the built-in field becomes significant or the onset of the efficiency droop~\cite{McSc2024}. Therefore, we adopt this same approach here for (Al,Ga)N QW systems.

Using the electronic structure obtained from the screened ETBM Hamiltonian, we determine the DOS and Urbach tail energies as follows. The hole energies from all configurations for each (Al,Ga)N QW system and carrier density are collated and grouped into energy bins. The bin sizes were adjusted for each system, to avoid for instance empty energy bins. Such an adjustment is required since the quantum confinement and thus the electronic structure is affected by changes in the well width as well as carrier density dependent screening effects. At the same time care must be taken not to chose bin sizes too large, as otherwise Urbach tails cannot be resolved.

   \begin{figure}[t!]
\includegraphics[width=\columnwidth]{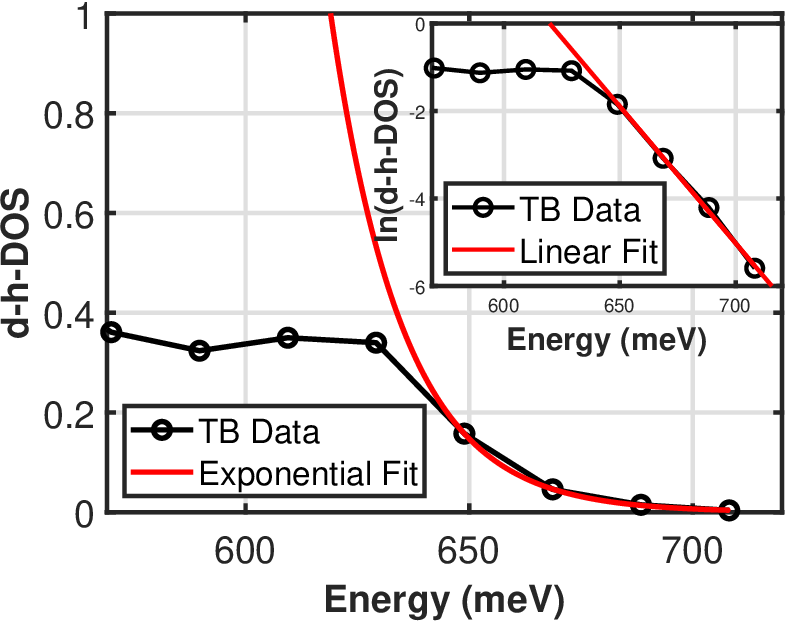}
\caption{Derivate of hole density of states, d-h-DOS, calculated for the 3.3 nm Al$_{0.48}$Ga$_{0.52}$N/Al$_{0.63}$Ga$_{0.37}$N quantum well at low carrier densities, i.e. in the absence of built-in field screening effects. Tight-binding (TB) data (black circles) and exponential fit to the Urbach tail (red line) are given. The inset shows the natural logarithm of the TB d-h-DOS data fitted with a linear function.}
\label{fig:dos}
\end{figure}

Taking the $L_w=3.3$ nm Al$_{0.48}$Ga$_{0.52}$N/Al$_{0.63}$Ga$_{0.37}$N QW as an example, Fig.~\ref{fig:dos} demonstrates how Urbach tail energies for the holes are determined. Instead of looking at the hole DOS (h-DOS) itself, we take the derivative of h-DOS (d-h-DOS) with respect to energy, $\epsilon$, numerically. The reason for this is that the modification to the DOS due to alloy disorder, the Urbach tail, can be described by an exponential tail, $\propto \exp{\left(-\frac{\epsilon}{E_u}\right)}$; in the `standard' DOS for an ideal QW (parabolic band) the h-DOS would exhibit a piecewise constant dependence of energy. By taking the derivative of h-DOS with respect to energy $\epsilon$ one can identify the Urbach tail region in the d-h-DOS, which is again $\propto \exp{\left(-\frac{\epsilon}{E_u}\right)}$. Thus this allows us to extract $E_u$ from the ETBM data. Figure~\ref{fig:dos} clearly shows the exponential tail in d-h-DOS. The inset figure displays the natural logarithm of the data, which reflects the expected straight line. From the slope of the line of best fit, $E_{u}$ is extracted. This method is employed below to determine $E_u$ for the considered QW systems at different carrier densities.

\section{Results}
\label{sec:results}

In this section we discuss Urbach tail energies and DOP of (Al,Ga)N QW systems. Special attention is paid to the impact of (i) Al content, (ii) well width and (iii) carrier density on the results. In Sec.~\ref{sec:Al48} we start with the lower Al content system, namely Al$_{0.48}$Ga$_{0.52}$N/Al$_{0.63}$Ga$_{0.37}$N QWs, before turning to the higher Al content structure, Al$_{0.75}$Ga$_{0.25}$N/Al$_{0.90}$Ga$_{0.10}$N, in Sec.~\ref{sec:Al75}.  

\subsection{Al$_{0.48}$Ga$_{0.52}$N/Al$_{0.63}$Ga$_{0.37}$N Quantum Well Systems}
\label{sec:Al48}

We analyze in section Sec.~\ref{sec:Al48_UT} the Urbach tail energies of Al$_{0.48}$Ga$_{0.52}$N/Al$_{0.63}$Ga$_{0.37}$N QWs of different widths. In Sec.~\ref{sec:Al48_DOP} the DOP is studied.

\subsubsection{Urbach Tails}
\label{sec:Al48_UT}

 \begin{figure}[t!]
\includegraphics[width=\columnwidth]{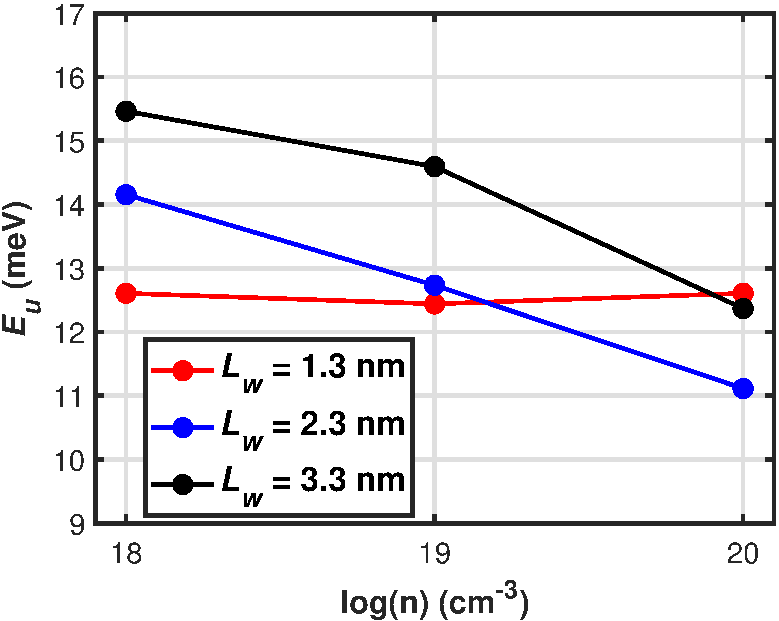}
\caption{Urbach tail energy $E_u$ for Al$_{0.48}$Ga$_{0.52}$N/Al$_{0.63}$Ga$_{0.37}$N quantum wells of width $L_w=1.3$ nm (red), $L_w=2.3$ nm (blue) and $L_w=3.3$ nm (black) at carrier densities of $n=1\times10^{18}$ cm$^{-3}$, $n=1\times10^{19}$ cm$^{-3}$ and $n=1\times10^{20}$ cm$^{-3}$.}
\label{fig:Eu48}
\end{figure}

 \begin{figure*}[t]
 \centering
\includegraphics[width=1.0\textwidth]{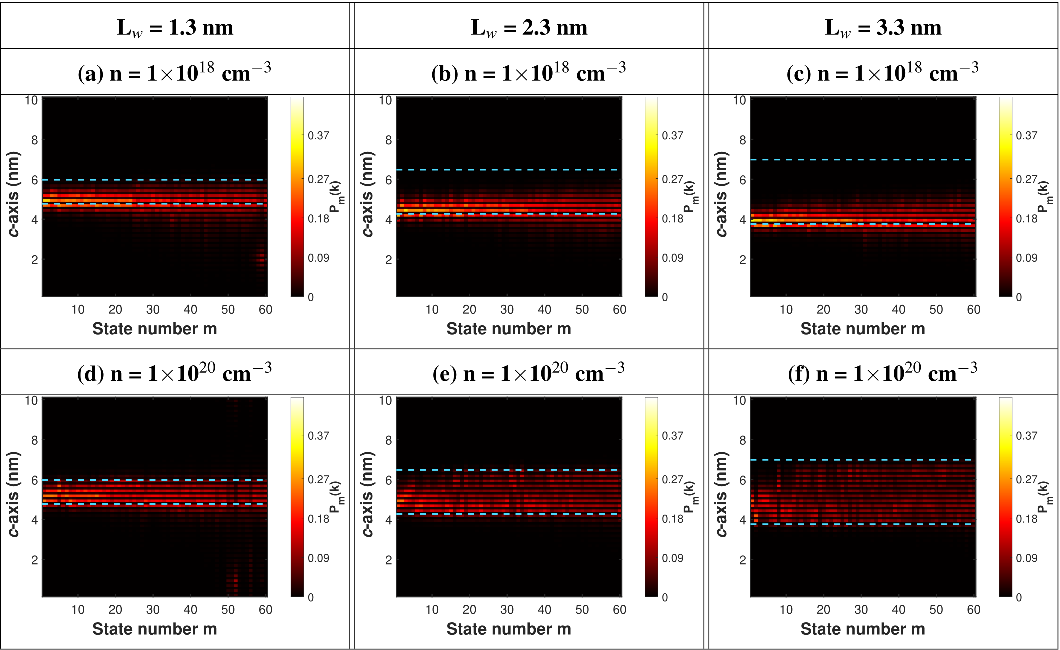}
\caption{Planar integrated probability densities of holes, $P_m(k)$, for an arbitrarily chosen alloy configuration of an Al$_{0.48}$Ga$_{0.52}$N/Al$_{0.63}$Ga$_{0.37}$N quantum well of width $L_w=1.3$ nm, $L_w=2.3$ nm and $L_w=3.3$ nm. The data are displayed at carrier densities of $n=1\times10^{18}$ cm$^3$ (a)-(c) and $n=1\times10^{20}$ cm$^3$ (d)-(f) for 60 hole states. Index $m$ refers to the single-particle hole state number, with $m=1$ being the ground state, while $k$ refers to the layer in the supercell along the wurtzite $c$-axis. The quantum well boundaries are indicated by the light blue dashed lines. The colourbar is kept the same between the different figures and is determined by the maximum $P_m(k)$ value found in the $L_w=3.3$ nm Al$_{0.75}$Ga$_{0.25}$N/Al$_{0.90}$Ga$_{0.10}$N quantum well system at a carrier density of $1\times10^{18}$ cm$^3$ discussed in Sec.~\ref{sec:Al75}.}
\label{fig:heat48}
\end{figure*}

Following the procedure outlined in Sec.~\ref{sec:QWModel}, Fig.~\ref{fig:Eu48} displays the extracted Urbach tail energies $E_{u}$ for Al$_{0.48}$Ga$_{0.52}$N/Al$_{0.63}$Ga$_{0.37}$N QWs of different widths $L_w$ as a function of carrier densities, $n$, in the well. In the low carrier density regime, $n=1 \times 10^{18}$ cm$^{-3}$, the Urbach tail energy increases with increasing $L_W$. This finding indicates that with increasing well width hole localization effects increase. 

Figure~\ref{fig:Eu48} also reveals that at least for the $L_w=2.3$ nm and $L_w=3.3$ nm system, the Urbach tail energy $E_u$ decreases with increasing carrier density. This finding indicates that alloy disorder induced carrier localization is amplified by the built-in field. However, in the well with $L_w=1.3$ nm we find that $E_u$ is approximately constant and may even slightly exceed the Urbach tail energy of the system with $L_w=2.3$ nm  and $L_w=3.3$ nm at high carrier densities.

To gain further insight into the interplay of carrier localization effects and built-in field, we discuss in the following the planar integrated probability density~\cite{ODoLu2021}
\begin{equation}
P_m(k) = \sum_{ij}\sum_{\alpha}|c^{\alpha,m}_{ijk}|^2 
 \label{eq:PLAN}
\end{equation}
which builds on the ETBM wave function $\psi_m$ expressed as:
\begin{equation}
\psi_m=\sum_{ijk} \sum_\alpha c^{\alpha,m}_{ijk}\phi^{\alpha}_{ijk}\,\, .
\end{equation}
The spatial $x$, $y$ and $z$ coordinates of the $N=81,920$ lattice sites in the supercell are labeled above by $i$, $j$ and $k$, respectively. The $sp^3$ ETBM basis states are denoted by $\phi^{\alpha}_{ijk}$ with $\alpha\in\left\{s, p_x , p_y, p_z\right\}$, and $c^{\alpha,m}_{ijk}$
being the expansion coefficient at each lattice site for the single particle hole state $m$. The expansion coefficients for a given state $m$ are obtained by diagonalising the ETBM Hamiltonian.

The quantity $P_m(k)$ gives the probability that the hole state $m$ is found in the layer specified by $k$ along the $c$-axis of the system. In Figs.~\ref{fig:heat48} (a)-(f), $P_m(k)$ is displayed for an arbitrarily chosen configuration for the three well widths considered. The data are plotted for low (\mbox{$n=1\times10^{18}$ cm$^{-3}$}) and high carrier density ($n=1\times10^{20}$ cm$^{-3}$) values; the colour bar is kept fixed for easier comparison between the different systems. 
In general, Figure~\ref{fig:heat48} (a)-(f) highlight carrier confinement effects along the $c$-axis arising from (i) the polarization field and (ii) the well width. However, information about in-plane localization effects is also encoded in $P_m(k)$: for a wave function that is delocalized in the $c$-plane and along the growth direction, $P_m(k)$ will not exhibit strong variations. Therefore, the absence of large variations in $P_m(k)$ for a given state is indicative of the absence of strong carrier localization effects both due to built-in field and alloy disorder, thus in the growth plane and along the $c$-axis.

Firstly, for the low carrier density of $n=1\times10^{18}$ cm$^{-3}$, Fig.~\ref{fig:heat48} (a)-(c) reveal the expected behavior, namely that the built-in polarization field confines the hole charge density to the lower QW barrier interface. Figures~\ref{fig:heat48} (a)-(c) also show that the impact of the built-in polarization field is less pronounced in the $L_w=1.3$ nm QW system, Fig.~\ref{fig:heat48} (a), as the potential drop is much smaller in this system when for instance compared to $L_w=3.3$ nm. In the $L_w=2.3$ nm, Fig.~\ref{fig:heat48} (b), and $L_w=3.3$ nm QWs, Fig.~\ref{fig:heat48} (c), the hole wave functions are clearly confined to much shorter lengths along the $\it{c}$-axis than the quantum confinement introduced by the respective QW width. The confinement lengths here may even be below or close to 1.3 nm.

Therefore, at low carrier densities, the hole wave function for the $L_w=2.3$ nm and $L_w=3.3$ nm QW experiences strong confining potentials which are driven by the large potential drop present in these systems. In the $L_w=1.3$ nm well, the quantum confinement due to the well width plays a larger role. Due to the strong confinement regime in all three cases, local fluctuations in Ga content can then lead to carrier localization effects, given the lower band gap of GaN, when compared to AlN, and the high effective hole masses in general~\cite{FiSc2022}.

With increasing carrier density in the well the built-in field is screened. Figures~\ref{fig:heat48} (d)-(f) display $P_m(k)$ now at the higher carrier density of $n=1\times10^{20}$ cm$^{-3}$. This screening effect is of secondary importance for the $L_w=1.3$ nm system as quantum confinement is to a large extent introduced by the well width and not the built-in potential. This is contrasted with the $L_w=2.3$ nm and $L_w=3.3$ nm wells, where the screening of the built-in field with increasing carrier density significantly impacts carrier localization. Although there are a few energy states with a high probability of being found in certain planes, overall the charge densities tend to become more delocalized along the $c$-axis within the QW region. As discussed above already, the absence of strong variations in $P_m(k)$ values indicates also the absence of strong in-plane carrier localization effects. The reduction in carrier localization effects with increasing carrier density explains the reduction in the Urbach tail energy shown in Fig.~\ref{fig:Eu48} for the wider wells. Our findings thus show that alloy disorder induced carrier localization effects in wider (Al,Ga)N QWs are significantly enhanced by polarization fields. With increasing carrier density these localization effects are strongly reduced with increasing built-in field screening effects because carrier confinement is mainly driven by the well width rather than polarization fields. However, in narrower wells this is not the case as the potential drop is small and even in the higher carrier density regime the quantum confinement is determined to a larger extent by the narrow well width. Thus, carrier localization may still be prevalent due to the reduced impact of the polarization field in narrower wells.

\subsubsection{Degree of Optical Polarization}
\label{sec:Al48_DOP}

To calculate the DOP, we follow previous work~\cite{FiSc2022} and focus on the orbital character of the occupied hole/valence states as a function of carrier density for a fixed temperature of \mbox{$T=300$ K}. We assume here that electron states are are predominantly $s$-like in character, which is confirmed by our ETBM calculations. Thus, it is sufficient to focus on the valence band structure. To do so we define the DOP, $\rho$, as follows:

\begin{equation}
\rho=\frac{\sum\limits_{\tilde{N}=1}^{150}\sum\limits_if(E_{i,\tilde{N}},T)(I^{i,\tilde{N}}_x+I^{i,\tilde{N}}_y-I^{i,\tilde{N}}_z)}{\sum\limits_{\tilde{N}=1}^{150}\sum\limits_if(E_{i,\tilde{N}},T)(I^{i,\tilde{N}}_x+I^{i,\tilde{N}}_y+I^{i,\tilde{N}}_z)}\,\, .
 \label{eq:DOLP}
\end{equation}

Here $\tilde{N}$ denotes the alloy configuration, $I_\alpha^{i,\tilde{N}}$ is the $p_x$ \mbox{($\alpha=x$)}, $p_y$ ($\alpha=y$) or $p_z$ ($\alpha=z$) orbital contribution in the $i$th QW state with energy $E_{i,\tilde{N}}$; $f(E_{i,\tilde{N}},T)$ denotes the Fermi function at temperature $T$, where the (quasi) Fermi level is defined with respect to the band edge energies and determined by the carrier density in the system. Using Eq.~(\ref{eq:DOLP}), a value of $\rho=1$ corresponds to TE polarized light emission from the well at a given carrier density at $T=300$ K, while $\rho=-1$ indicates TM polarized emission. 
The above approach allows us to gain insight into the light polarization properties in general. It does not provide insight into the relative strength of TM to TE polarized transition, i.e. it does not provide a full emission spectrum. The above can be understood as a necessary condition for TE or TM polarized light emission: in the absence of e.g. a $p_z$-like orbital character in a hole state, TM polarized light emission is suppressed/absent. Future studies may target polarization resolved emission spectra~\cite{ScTa2015}, but to shed light onto how built-in polarization fields, alloy disorder and well width affect the DOP, Eq.~(\ref{eq:DOLP}) provides a good starting point for such an investigation.

\begin{figure}
\centering
\includegraphics[width=\columnwidth]{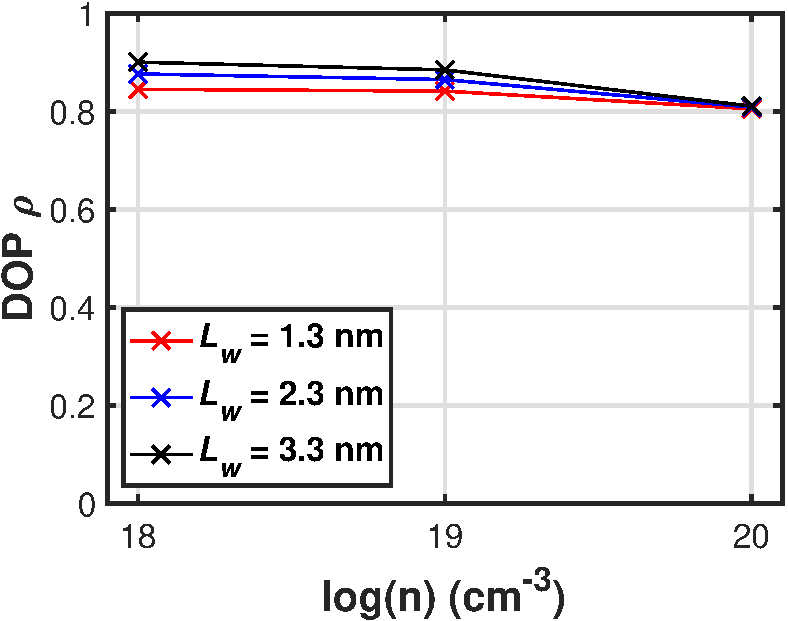}
\caption{Averaged degree of optical polarization, DOP $\rho$, over 150 configurations at a temperature of 300 K for Al$_{0.48}$Ga$_{0.52}$N/Al$_{0.63}$Ga$_{0.37}$N quantum wells of width $L_w=1.3$ nm (red), $L_w=2.3$ nm (blue) and $L_w=3.3$ nm (black) as a function of the carrier density $n$.}
\label{fig:dop48}
\end{figure}

Figure~\ref{fig:dop48} displays the DOP for the different well widths considered, which reveals that the emission of the Al$_{0.48}$Ga$_{0.52}$N QW system studied here is strongly TE polarized and that well width and carrier density only slightly affect the DOP. These findings are in line with experimental data from the literature~\cite{SuZi2020,ReGu2015APL}. Looking at the data displayed in Fig.~\ref{fig:dop48} in more detail, one observes that with increasing well width $L_w$ the DOP slightly increases, especially when considering a low carrier density $n=1\times10^{18}$ cm$^{-3}$. Moreover, and independent of $L_w$, we find that with increasing carrier density the DOP slightly decreases. We attribute these effects to changes in the electrostatic polarization field. As discussed in the previous section, the larger potential drop in the wider wells leads to stronger wave function localization effects at the lower QW barrier interface, cf. Fig.~\ref{fig:heat48}. Given that $p_z$-like basis states exhibit a lower effective mass along the $c$-axis, see discussion in Sec.~\ref{sec:VB}, one expects that states exhibiting a large $p_z$-orbital contribution are pushed to lower energies and thus do not contribute significantly to the DOP. This explains why the DOP increases slightly with increasing well width. On the other hand, with increasing carrier density, the polarization field is screened and so are carrier localization effects, especially for wider wells. Thus, states with $p_z$-orbital contribution may be found energetically closer to the VBE. Based on Eq.~(\ref{eq:DOLP}), if states with higher $p_z$-orbital character are being populated, the DOP reduces, explaining thus the slight decrease in DOP with increasing carrier density. 

Overall, our calculations show that the DOP in Al$_{0.48}$Ga$_{0.52}$N/Al$_{0.63}$Ga$_{0.37}$N QWs is largely determined by macroscopic effects such as strain and polarization field effects. As we will see in the following, this is not necessarily the case at higher Al contents in the well and barrier.    

\begin{figure}[b!]
\includegraphics[width=\columnwidth]{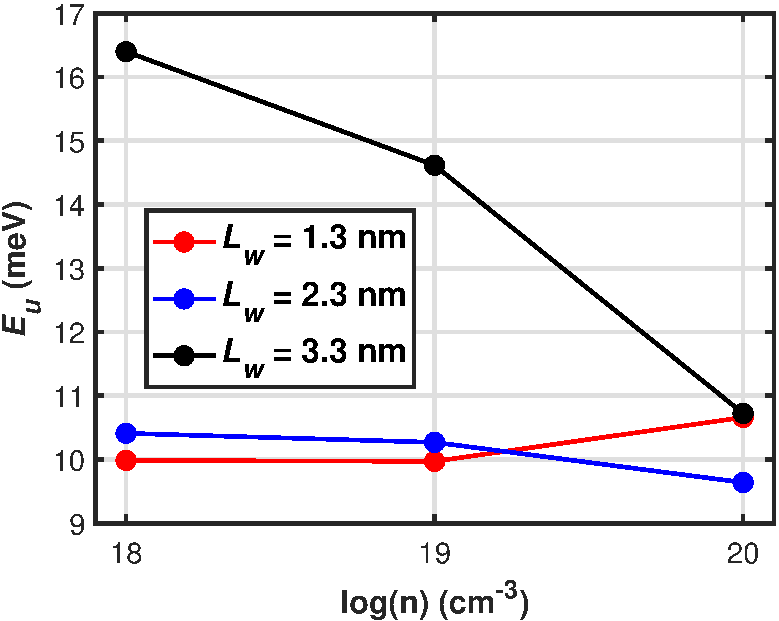}
\caption{Urbach tail energies, $E_u$, for Al$_{0.75}$Ga$_{0.25}$N/Al$_{0.90}$Ga$_{0.10}$N quantum wells of width $L_w=1.3$ nm (red), $L_w=2.3$ nm (blue) and $L_w=3.3$ nm (black) at carrier densities of $n=1\times10^{18}$ cm$^{-3}$, $n=1\times10^{19}$ cm$^{-3}$ and $n=1\times10^{20}$ cm$^{-3}$.}
\label{fig:Eu75}
\end{figure}

\begin{figure*}
\includegraphics[width=1.0\textwidth]{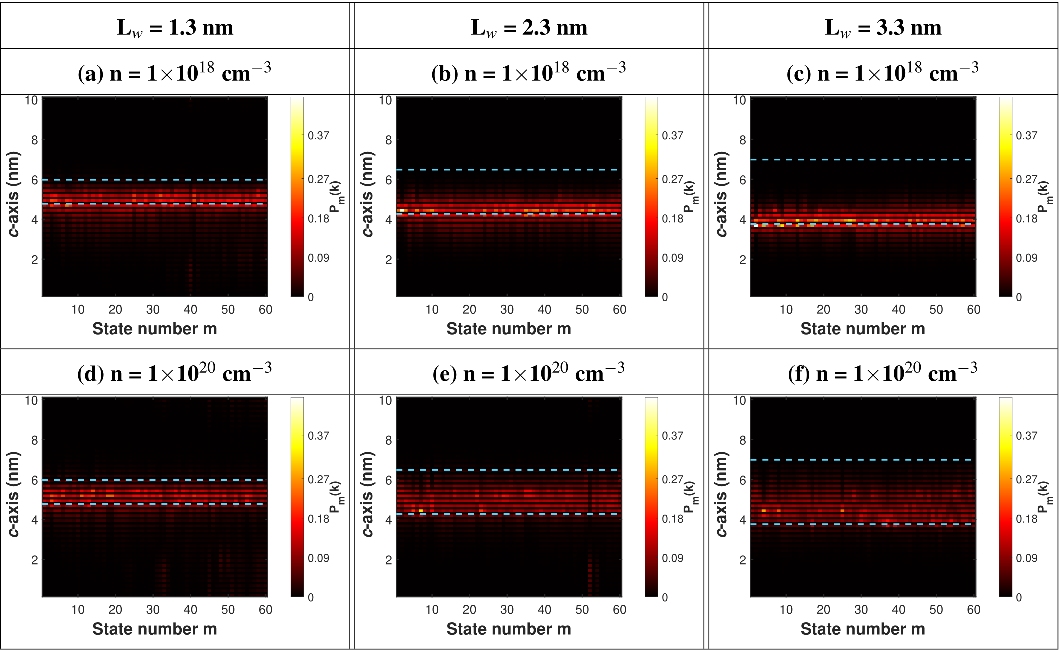}
\caption{Planar integrated probability densities of holes, $P_m(k)$, for an arbitrarily chosen configuration of an Al$_{0.75}$Ga$_{0.25}$N/Al$_{0.90}$Ga$_{0.10}$N quantum well of width $L_w=1.3$ nm, $L_w=2.3$ nm and $L_w=3.3$ nm. The data are displayed at carrier densities of $n=1\times10^{18}$ cm$^3$ (a)-(c) and $n=1\times10^{20}$ cm$^{-3}$ (d)-(f) for 60 states. Index $m$ refers to the single-particle hole state number, with $m=1$ being the ground state, while $k$ refers to the layer in the supercell along the wurtzite $c$-axis. The quantum well boundaries are indicated by the light blue dashed lines. $P_m(k)$ has been plotted to the maximum $P_m(k)$ value found in the $L_w=3.3$ nm Al$_{0.75}$Ga$_{0.25}$N/Al$_{0.90}$Ga$_{0.10}$N QW system at a carrier density of $1\times10^{18}$ cm$^3$.}
\label{fig:heat75}
\end{figure*}

\subsection{Al$_{0.75}$Ga$_{0.25}$N/Al$_{0.90}$Ga$_{0.10}$N Quantum Well System}
\label{sec:Al75}
Having discussed Urbach tail energies and DOP for the lower Al content wells in the previous section, we turn now and investigate these quantities for the higher Al content structures with Al$_{0.75}$Ga$_{0.25}$N/Al$_{0.90}$Ga$_{0.10}$N. As highlighted in Sec.~\ref{sec:QWModel}, such a system would be considered when targeting emitters in the deep UV spectral range and where competition between TE and TM polarization is important for the LEE. We start again with Urbach tail energies, Sec.~\ref{subsec:UE_Al75}, before turning to the DOP, Sec.~\ref{subsec:DOP_Al75}.  

\subsubsection{Urbach Tails}
\label{subsec:UE_Al75}

Figure~\ref{fig:Eu75} depicts the Urbach tail energies for Al$_{0.75}$Ga$_{0.25}$N/Al$_{0.90}$Ga$_{0.10}$N QWs of $L_w=1.3$ nm, $L_w=2.3$ nm, and $L_w=3.3$ nm, as a function of the carrier density $n$.  

For the $L_w=3.3$ nm QW system, $E_u$ decreases with increasing carrier density. In contrast, $E_{u}$ exhibits a much weaker carrier density dependence in the $L_w=1.3$ nm and $L_w=2.3$ nm wells; also the $E_u$ values for $L_w=1.3$ nm and $L_w=2.3$ nm systems are (much) smaller in magnitude when compared to the $L_w=3.3$ nm Al$_{0.75}$Ga$_{0.25}$N QW or equivalent Al$_{0.48}$Ga$_{0.52}$N wells, cf. Fig.~\ref{fig:Eu48}. Furthermore, Fig.~\ref{fig:Eu75} shows that the Urbach tail energies in the $L_w=1.3$ nm and $L_w=2.3$ nm wells behave differently with increasing $n$: while $E_u$ slightly increases with increasing $n$ for $L_w=1.3$ nm, for $L_w=2.3$ nm $E_u$ slightly decreases.

\begin{figure*}
\includegraphics[width=1.0\textwidth]{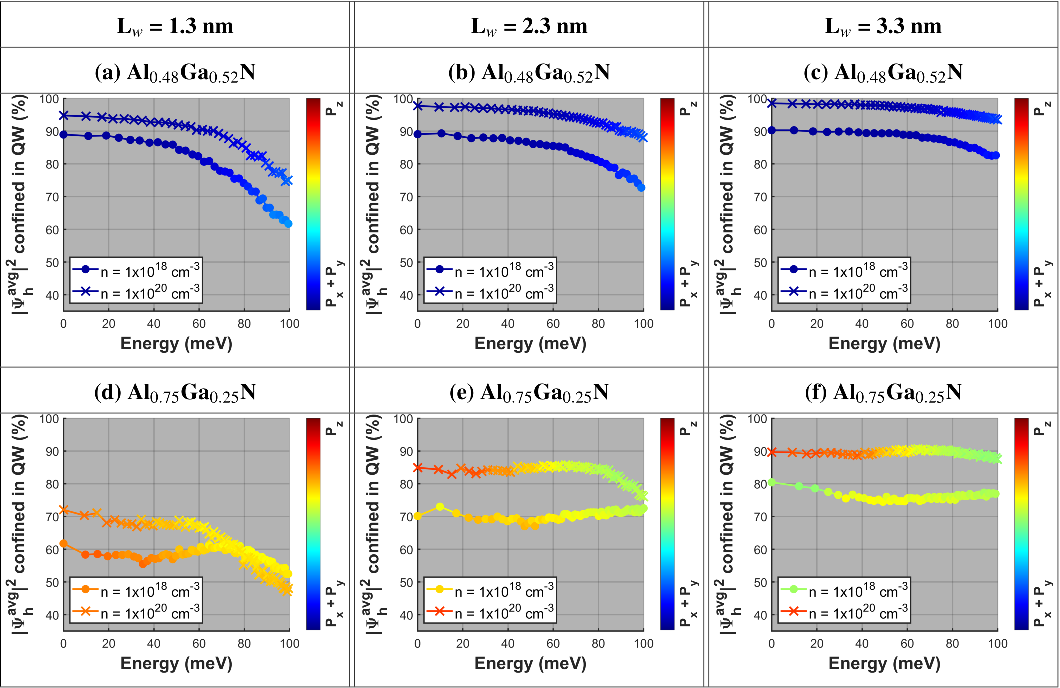}
\caption{Average probability of hole charge density, $|\Psi^{avg}_h|^2$, confined inside the (Al,Ga)N quantum wells as a function of energy. The data are averaged over the 150 configurations. The average state energy is plotted with respect to the averaged ground state energy of each quantum well system. The Al$_{0.48}$Ga$_{0.52}$N/Al$_{0.63}$Ga$_{0.37}$N wells are shown for (a) $L_w=1.3$ nm, (b) $L_w=2.3$ nm, and (c) $L_w=3.3$ nm; the data for Al$_{0.75}$Ga$_{0.25}$N/Al$_{0.90}$Ga$_{0.10}$N wells is displayed in (d)-(f). For each QW system $|\Psi^{avg}_h|^2$ is plotted for two carrier densities: $n=1\times10^{18}$ cm$^{-3}$ (circles) and $n=1\times10^{20}$ cm$^{-3}$ (crosses). The color coding gives the averaged orbital contribution of each energy level (red: $p_z$; blue: $p_x+p_y$).}
\label{fig:QW_conf}
\end{figure*}

To shed more light onto these findings, Fig.~\ref{fig:heat75} depicts the planar integrated hole probability densities $P_m(k)$ for the Al$_{0.75}$Ga$_{0.25}$N QWs considered.
Looking at the \mbox{$L_w=3.3$ nm} QW system first, we find a behaviour similar to the Al$_{0.48}$Ga$_{0.52}$N wells discussed in Sec.~\ref{sec:Al48}. In the low carrier density regime of $n=1\times10^{18}$ cm$^{-3}$, the built-in potential leads to a strong localization of the hole wave functions at the QW barrier interface, cf. Fig~\ref{fig:heat75} (c). When the carrier density reaches $n=1\times10^{20}$ cm$^{-3}$, the built-in field is screened and carrier localization effects are reduced, cf. Fig~\ref{fig:heat75} (f). In the high carrier density regime, there are only a few extremely high $P_m(k)$ values, indicating thus that alloy disorder induced carrier localization effects are weaker in this regime. Overall, these findings explain the reduction in $E_u$ with increasing $n$.

Turning to the $L_w=1.3$ nm QW and comparing $P_m(k)$ for $n=1\times10^{18}$ cm$^{-3}$, cf. Fig~\ref{fig:heat75} (a), and $n=1\times10^{20}$ cm$^{-3}$, cf. Fig~\ref{fig:heat75} (d), one observes that carrier localization is very different for these systems, especially for the states near the VBE, thus close to the ground state $m=1$. While in the high carrier density regime, $n=1\times10^{20}$ cm$^{-3}$, the hole charge densities are localized \emph{inside} the well, for the low carrier density case, $n=1\times10^{18}$ cm$^{-3}$, the charge density spreads noticeably into the barrier material. This indicates a weaker carrier confinement in the $n=1\times10^{18}$ cm$^{-3}$ case when compared to $n=1\times10^{20}$ cm$^{-3}$. This means that screening the built-in field results in improved carrier confinement for the narrow well. The stronger confinement in the well leads to an increase in the Urbach tail energy $E_u$ when the carrier density increases to $n=1\times10^{20}$ cm$^{-3}$.

The $L_w=2.3$ nm QW system behaves similar to $L_w=1.3$ nm in the low carrier density regime of $n=1\times10^{18}$ cm$^{-3}$, meaning that the hole wave functions also leak into the barrier material. Thus the Urbach tail energies at $n=1\times10^{18}$ cm$^{-3}$ for $L_w=1.3$ nm and $L_w=2.3$ nm are expected to be similar in magnitude, consistent with Fig.~\ref{fig:Eu75}. When screening the polarization field by increasing the carrier density in the well to $n=1\times10^{20}$ cm$^{-3}$, cf. Fig.~\ref{fig:heat75} (e), the hole wave functions start to localize inside the well again. However, in comparison to the $L_w=1.3$ nm system, the potential drop is larger in the $L_w=2.3$ nm QW system and as such screening of the built-in field becomes more important. At the highest carrier density, wave functions are now mainly confined by the well width as the potential drop is strongly reduced due to built-in field screening effects. Therefore one may expect to see a reduction in the Urbach tail energies in the $L_w=2.3$ nm system $n=1\times10^{20}$ cm$^{-3}$ when compared to the $L_w=1.3$ nm system, cf. Fig.~\ref{fig:Eu75}, due to the hole wave function extending along a greater well width and do not sample alloy disorder of a few monolayers as in the case of the strong built-in field at low carrier densities. In general, a reduction of the Urbach tail energy can be attribute to a delocalization of the hole wave functions. This is supported by Fig.~\ref{fig:heat75} in the sense that (i) the wave functions spread over a larger region of the QW with increasing $n$ and (ii) the absence of large $P_m(k)$ values on only a few layers $k$.  

\begin{figure}[t]
\centering
\includegraphics[width=\columnwidth]{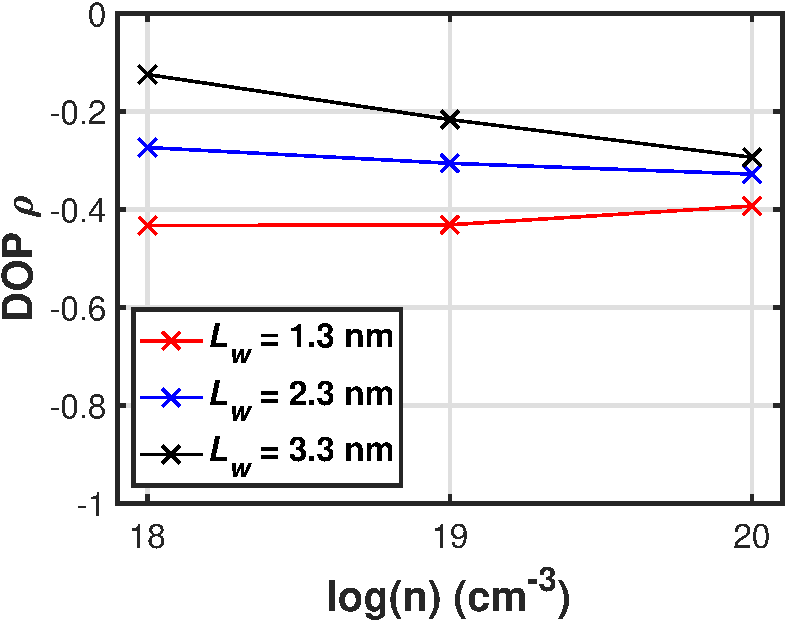}
\caption{Averaged degree of optical polarization, DOP $\rho$, over 150 configurations at a temperature of 300 K for Al$_{0.75}$Ga$_{0.25}$N/Al$_{0.90}$Ga$_{0.10}$N quantum wells of width $L_w=1.3$ nm (red), $L_w=2.3$ nm (blue) and $L_w=3.3$ nm (black) as a function of carrier density $n$.}
\label{fig:dop75}
\end{figure}

To further highlight changes in the carrier confinement and its impact of wave function localization effects, Fig.~\ref{fig:QW_conf} shows the probability of finding the hole charge density inside the QW, $|\Psi^{avg}_h|^2$, as a function of energy $E_i^{avg}$. The data are averaged over all 150 configurations and the energy is shifted to the average ground state energy $E^{avg}_1$ for each QW system. Here, a higher energy means further away from the VBE. For comparison, Fig.~\ref{fig:QW_conf} (a)-(c) displays $|\Psi^{avg}_h|^2$ for Al$_{0.48}$Ga$_{0.52}$N wells, while Fig.~\ref{fig:QW_conf} (d)-(f) depicts the data for the Al$_{0.75}$Ga$_{0.25}$N wells, at carrier densities of $1\times10^{18}$ cm$^{-3}$ and $1\times10^{20}$ cm$^{-3}$. Each energy level has been colour coded to visualise the average orbital character. Figure~\ref{fig:QW_conf} reveals that with increasing carrier density the wave function confinement inside the QW is increased due to screening of the polarization field. However, we note also that even though the Al contrast between well and barrier is the same in Al$_{0.75}$Ga$_{0.25}$N/Al$_{0.90}$Ga$_{0.10}$N and Al$_{0.48}$Ga$_{0.52}$N/Al$_{0.63}$Ga$_{0.37}$N systems, the confinement is in general weaker in the high Al content system, cf. Fig.~\ref{fig:QW_conf}. Narrow QWs with high Al contents are widely used in deep UV LEDs~\cite{HoKu2024APL,ReGu2015APL}. These structures often suffer from low EQEs. Our calculations provide a possible reason for this behavior, if the wave functions are weakly confined, recombination at defects may be more likely, due to a higher probability of the hole charge densities at a defect site, assuming a delocalized electron wave function. Thus our findings suggest that using wider wells for (Al,Ga)N-based deep UV LEDs can be beneficial for the device performance. As expected from our discussion in Sec.~\ref{sec:VB}, Fig.~\ref{fig:QW_conf} also confirms that the orbital character of the hole states is sensitive to QW width and carrier density. Thus the DOP will also be impacted by these factors. In the following section we discuss the impact of well width and carrier density on the DOP in Al$_{0.75}$Ga$_{0.25}$N/Al$_{0.90}$Ga$_{0.10}$N QWs in more detail.

\subsubsection{Degree of Optical Polarization}
\label{subsec:DOP_Al75}

Before turning to the DOP in Al$_{0.75}$Ga$_{0.25}$N/Al$_{0.90}$Ga$_{0.10}$N QWs we revisit briefly the Al$_{0.48}$Ga$_{0.52}$N wells. 
The colour coding in Fig.~\ref{fig:QW_conf} (a)-(c) reveals that in the Al$_{0.48}$Ga$_{0.52}$N QWs the hole states in an energy range of 100 meV below the VBE are predominately $p_x$ and $p_y$-like in orbital character. Building on our definition for the DOP $\rho$, Eq.~(\ref{eq:DOLP}), this confirms the earlier finding that independent of well width or carrier density, light emitted from the Al$_{0.48}$Ga$_{0.52}$N QWs will be predominately TE polarized. 

Turning in a second step to the Al$_{0.75}$Ga$_{0.25}$N systems, Fig.~\ref{fig:QW_conf} (d)-(f), one finds a more complex behavior. Firstly one observes that the orbital character of the states varies with energy. Independent of the well width or carrier density, states near the VBE are a mixture of $p_x$, $p_y$ and $p_z$-like orbitals, in contrast to the Al$_{0.48}$Ga$_{0.52}$N wells. Looking at the low carrier density of $n=1\times10^{18}$ cm$^{-3}$ results for the three different wells, Fig.~\ref{fig:QW_conf} (d)-(f), indicate that with increasing well width the $p_z$-orbital contribution in the states energetically closest to VBE, i.e. the zero of energy, is reduced. This is indicative of an increasing DOP with increasing well width at low carrier densities. One may expect the opposite behavior, given that the quantum confinement due to the increasing well width is reduced and that the states with predominately $p_z$ orbital character may shift to energies energetically closer to the VBE; this expectation is based on the fact that $p_z$-like states exhibit a lower effective mass along the $c$-axis.  
However, with increasing well width, the potential drop across the well also increases and thus increases the carrier confinement. This is also reflected in the increase of $|\Psi^{avg}_h|^2$ with increasing well width at the fixed carrier density of $n=1\times10^{18}$ cm$^{-3}$. Therefore, due to the lower effective mass of $p_z$-like states along the $c$-axis growth direction, stronger confinement arising from the built-in potential will shift $p_z$-like states energetically away from the VBE.

With increasing carrier density the built-in field is screened and the confinement effect introduced by this field is reduced. Consequently, at a high carrier density of \mbox{$n=1\times10^{20}$ cm$^{-3}$}, the $p_z$ orbital character near the VBE increases with increasing well width $L_w$ as Fig.~\ref{fig:QW_conf} (d)-(f) reveals. Thus, at high carrier densities, the DOP is expected to decrease with increasing well width $L_w$.

Equipped with this insight, we use Eq.~(\ref{eq:DOLP}) to study the DOP in the Al$_{0.75}$Ga$_{0.25}$N/Al$_{0.90}$Ga$_{0.10}$N QWs as a function of carrier density at a fixed temperature of $T=300$ K. The calculated DOP values are shown in Fig.~\ref{fig:dop75}. The data indicate that the emission is indeed noticeable TM polarized for this high Al content system, which is in contrast to the Al$_{0.48}$Ga$_{0.52}$N/Al$_{0.63}$Ga$_{0.37}$N wells investigated in Sec.~\ref{sec:Al48} but in line with similar experimental observations in the literature~\cite{IbLe2024}. 

Looking at the DOP as a function of carrier density $n$ for a fixed well width, Fig.~\ref{fig:dop75} shows that for the narrow well width of $L_w=1.3$ nm the DOP increases with increasing carrier density. We attribute this to the effect that at low carrier densities the hole wave functions are much weaker bound in the $L_w=1.3$ nm QW system when compared to the high carrier density regime, which is supported by the data in Fig~\ref{fig:QW_conf}. In the wider wells, $L_w=2.3$ nm and $L_w=3.3$ nm, the DOP decreases with increasing carrier density. The observed behaviour stems from changes in the confining potential discussed above, originating from the screening of the built-in field with increasing carrier density. Also, given that these screening effects are more pronounced in the wider well, the DOP changes with increasing carrier density $n$ more strongly in the $L_w=3.3$ nm system when compared to $L_w=2.3$ nm. However, even at high carrier densities of $n=1\times10^{20}$ cm$^{-3}$ a slightly larger DOP is observed in the $L_w=3.3$ nm system when compared to $L_w=2.3$ nm and especially $L_w=1.3$ nm. At lower carrier densities this difference is even larger, and favors the wider wells even more for a higher DOP value. 

While narrow wells, e.g. $L_w=1$ nm or $L_w=1.5$ nm, are often employed in deep UV light sources~\cite{HoKu2024APL,ReGu2015APL}, our analysis indicates that wider wells can be beneficial for these emitters for several reasons. Firstly, as shown above, in the higher Al content regime, carrier confinement in narrow wells can be poor. While this may be an advantage for carrier transport in multi-QW (MQW) systems, these systems can also be prone to defect related recombination processes. Secondly,  wider wells are promising candidates in order to lower the steady state carrier density inside the well which may lead to a reduction in for instance Auger-Meitner related recombination processes~\cite{MuTu2019}. Based on our DOP calculations, wider wells with lower carrier densities will also lead to a higher DOP value when compared to the same carrier density in a narrower well. However, it should also be noted that in general it is expected that the radiative recombination rate is reduced in wider wells when compared to narrower wells due to the quantum confined Stark effect. Further studies are now required to analyse the competition between radiative, non-radiative and DOP in deep UV light emitters.

\section{Summary and Conclusion}
\label{sec:Concl}

In this work we have analyzed the DOP in (Al,Ga)N QWs operating in the UV-C spectral range by means of an atomistic multi-band electronic structure model. Our model accounts for alloy disorder induced band mixing effects which is neglected in widely available continuum based simulation frameworks. In our study special attention has been paid on the impact of (i) Al content, (ii) carrier density and (iii) well width on the DOP.

We have considered two (Al,Ga)N based QW well systems which differ in the Al content in the well and barrier. However, for consistent comparison, the Al contrast between well and barrier is kept constant in our investigations. The considered QW systems, Al$_{0.48}$Ga$_{0.52}$N/Al$_{0.63}$Ga$_{0.37}$N and Al$_{0.75}$Ga$_{0.25}$N/Al$_{0.90}$Ga$_{0.10}$N wells, are representative for structures found in the literature to achieve light emission in the deep UV-C spectral range, Al$_{0.75}$Ga$_{0.25}$N/Al$_{0.90}$Ga$_{0.10}$N, and the longer wavelengths end of the UV-C window, Al$_{0.48}$Ga$_{0.52}$N/Al$_{0.63}$Ga$_{0.37}$N. Even though the contrast in Al content is kept constant between these two systems, our calculations show that carrier confinement, especially for narrower wells, e.g. $L_w=1.3$ nm, is much weaker in the high Al content system. This effect may be beneficial for carrier transport in deep UV light emitters with high Al content, however, it can lead also to the situation that these emitters are more prone to defect related recombination since the wave functions can significantly spread into the barrier material.

In terms of the DOP, our calculations show that in general wider wells lead to a higher DOP value. In the lower Al content system, Al$_{0.48}$Ga$_{0.52}$N/Al$_{0.63}$Ga$_{0.37}$N wells, this observation is of secondary importance as the light is predominately transverse electric (TE) polarized, which can enable a high light extraction efficiency (LEE). This finding of predominantly TE polarized emission is independent of the well widths and carrier densities studied here. The situation is more complicated in the higher Al content Al$_{0.75}$Ga$_{0.25}$N/Al$_{0.90}$Ga$_{0.10}$N QW system. In general, with increasing well width the DOP increases, so that emission is tending towards TE polarization. We observe in our study that this effect is more pronounced at lower carrier densities, e.g. $n=1\times10^{18}$ cm$^{-3}$. However, even at high carrier densities of $n=1\times10^{20}$ cm$^{-3}$, an Al$_{0.75}$Ga$_{0.25}$N/Al$_{0.90}$Ga$_{0.10}$N QW of width $L_w=3.3$ nm still exhibits a slightly higher DOP when compared to the same system with a width of $L_w=1.3$ nm. 

Overall, for (Al,Ga)N-based light emitters operating in the deep UV spectral range our calculations indicate that, in terms of carrier confinement and light polarization characteristics, wider wells are advantageous. However, it should also be noted that wider wells may lead to a reduction in the radiative recombination rate which may offset the benefits of wider wells in terms of stronger carrier confinement in the well and an increased DOP. Therefore, future studies may target a detailed analysis of balancing these competing factors as a function of the well width in deep UV light emitters to improve their efficiencies.

\begin{acknowledgments}
This work received funding from the Sustainable Energy Authority of Ireland and the Science Foundation Ireland (Nos. 17/CDA/4789, 12/RC/2276 P2 and 21/FFP-A/9014), 
Leibniz competition 2022 (UVSimTec, K415/2021). 

The data that support the findings of this study are available from the corresponding author upon reasonable request.
\end{acknowledgments}

\providecommand{\noopsort}[1]{}\providecommand{\singleletter}[1]{#1}%

\end{document}